\documentclass{iopart}
\usepackage{iopams}
\usepackage{epsfig}
 
\begin{document} 
\jl{3}
 
 
\letter{Conductance anomalies in quantum wires} 
\author{T. Rejec$^a$ and A. Ram\v sak$^{a,b}$} 
\address{$^a$J. Stefan Institute,  SI-1000 Ljubljana, Slovenia \\ 
$^b$Faculty of Mathematics and Physics, University of Ljubljana, 
SI-1000 Ljubljana, Slovenia 
} 
\vskip -.5 cm
\author{J.H. Jefferson} 
\address{DERA, St. Andrews Road, Great Malvern, 
Worcestershire WR14 3PS, England 
}
\vskip 0.3 cm
\address{\hskip 4 cmFebruary 21, 2000}
                
\begin{abstract} 
We study the conductance threshold of clean nearly straight quantum
wires in which an electron is bound. We show that such a
system exhibits spin-dependent conductance structures on the rising
edge to the first conductance plateau, one near $0.25(2e^2/h)$,
related to a singlet resonance, and one near $0.75(2e^2/h)$, related
to a triplet resonance.  As a quantitative example we solve exactly 
the scattering problem for two-electrons in a wire with planar
geometry and a weak bulge. From the scattering matrix we determine
conductance via the Landauer-B\" uttiker formalism.  The conductance
anomalies are  robust and survive to temperatures
of a few degrees.  With increasing in-plane magnetic field the
conductance exhibits a plateau at $e^2/h$, consistent with
recent experiments.
\end{abstract} 
   
\noindent 
Following the pioneering work in Refs. \cite{wees88,wharam88} many
groups have  now observed conductance steps in various types of quantum
wire. These first experiments were performed on gated two-dimensional
electron gas (2DEG) structures, while similar behaviour of conductance
are shown in ``hard-confined'' quantum wire
structures, produced by cleaved edge over-growth
\cite{yacoby96},  epitaxial growth on ridges 
\cite{ramvall97},  heteroepitaxial growth on ``v''-groove 
surfaces \cite{walther92} and most recently in
GaAs/Al$_{\delta}$Ga$_{1-\delta}$As 
narrow ``v''-groove \cite{kaufman99} and low-disorder \cite{kane98}
quantum wires.

These experiments strongly support the idea of ballistic conductance
in quantum wires and are in surprising agreement with now the standard
Landauer-B\"{u}ttiker formalism \cite{landauer57,buttiker86}
neglecting electron interactions \cite{houten92}. However, there are
certain anomalies, some of which are believed to be related to
electron-electron interactions and appear to be spin-dependent. In
particular, already in early experiments a structure is seen in the
rising edge of the conductance curve \cite{wees88}, starting at around
$0.7 G_0$ with $G_0=2e^{2}/h$ and merging with the first conductance
plateau with increasing energy. Under increasing in-plane magnetic field, the
structure moves down, eventually merging with a new conductance
plateau at $e^{2}/h$ in very high fields
\cite{thomas96,thomas98}. Theoretically this anomaly has not been
adequately explained, despite several scenarios, including
spin-polarised sub-bands \cite{fasol94}, conductance suppression in a
Luttinger liquid with repulsive interaction and disorder
\cite{maslov95} or local spin-polarised density-functional theory
\cite{wang98}.  
Recently we have shown that these conductance anomalies near $0.7G_0$
and $0.25G_0$ are consistent with an electron being weakly bound in
wires of circular cross section, giving rise to spin-dependent
scattering resonances \cite{rejec99}. 

In this letter we extend our previous work to planar quantum wires
with rectangular cross section and also analyse the effects of an
external in-plane magnetic field.  We consider here, as an example, a
small fluctuation in thickness of the wire in some region giving rise
to a weak bulge. Such a system may be regarded as an ``open quantum
dot'' in which one electron is bound and inhibits the transport of
conduction electrons via Coulomb repulsion.  The problem is analogous
to treating the collision of an electron with a hydrogen atom as,
e.g., described in Ref.~\cite{landauQM} and studied 70 years ago by
J.R. Oppenheimer and N.F. Mott \cite{oppenheimer28}. The conductance
is obtained from the transmission probabilities for individual
channels via the usual Landauer-B\"{u}ttiker formalism
\cite{landauer57,buttiker86}. In the present two-electron problem, the
relevant channels are singlet and triplet, with transmission
amplitudes $t_{\rm s}$ and $t_{\rm t}$, respectively and corresponding
transmission probabilities ${\cal T}_{\rm s}(E)= |t_{\rm s}|^2$ and
${\cal T}_{\rm t}(E)=|t_{\rm t}|^2$. The transmission amplitudes for
particular spin configurations of the target (bound electron) and
scattered electron are further expressed in terms of $t_{\rm s}$ and
$t_{\rm t}$ as
$t_{\uparrow \uparrow \rightarrow \uparrow \uparrow }(E) = 
t_{\rm{t}}$, 
$t_{\downarrow \uparrow \rightarrow \downarrow \uparrow }(E)  = 
\frac{1}{2}(t_{\rm{s}}
+t_{\rm{t}})$, and 
$t_{\downarrow \uparrow \rightarrow \uparrow \downarrow }(E) = 
\frac{1}{2}(t_{\rm{s}}
-t_{\rm{t}})$, 
where, for example, $t_{\downarrow \uparrow \rightarrow \uparrow
\downarrow }(E)$ is the transition amplitude from
$[\downarrow,\uparrow]$ to $[\uparrow, \downarrow]$ spin states of the
$[$scattered,bound$]$ electron.  The conductance for unpolarised
electrons is then the {\it average} over initial and {\it sum} over
final configurations, $G(E)=G_0 {\cal T}(E)$, where
\begin{eqnarray}
\fl{\cal T}(E)=\frac{1}{4}\left( 
\left| t_{\uparrow \uparrow \rightarrow 
\uparrow \uparrow }\right| ^{2}+\left| t_{\downarrow \uparrow 
\rightarrow \downarrow \uparrow }\right| ^{2
}+\left| t_{\downarrow \uparrow \rightarrow \uparrow \downarrow }
\right| ^{2}+\left| t_{\downarrow \downarrow \rightarrow \downarrow 
\downarrow }\right| ^{2}+\left| t_{\uparrow \downarrow \rightarrow 
\uparrow \downarrow }\right| ^{2}+\left| t_{\uparrow \downarrow 
\rightarrow \downarrow \uparrow }\right| ^{2}\right)\nonumber\\
\lo=\frac{1}{4}{\cal T}_{\rm{s}}(E)+\frac{3}{4}
{\cal T}_{\rm{t}}(E).
\label{st}
\end{eqnarray}         
At finite temperatures the conductance is calculated using a generalised
Landauer-B\" uttiker formula \cite{bagwell89} 
\begin{equation}
G(E)=G_{0}\int{\cal T}(\epsilon )\left[ -\frac{\partial
f(\epsilon \!-\!E,T)}{\partial \epsilon }\right] \;{\rm d}\epsilon ,
\label{landauerT}
\end{equation}
where $f(\epsilon ,T)=[1+\exp (\epsilon /k_{{\rm B}}T)]^{-1}$ is the usual
Fermi function. 

For simplicity, quantitative treatment of quantum wires is restricted
to the geometry shown in Fig.~1, with confinement in the $x$- and
$y$-direction and electron propagates in the $z$-direction. This is
similar to wires produced in ``v''-grooves as reported by Kaufman {\it
et al.} \cite{kaufman99} with thicknesses in the range $10$ to
$20$~nm.  The wire shape under consideration is symmetric around the
$z$ axis with constant potential, $V(x,y,z)=0$ within a boundary shown
in Fig.~1, and confining potential $V_{0}>0$ elsewhere. As shown in
Fig.~1 the wire has thickness $a_3$ and a single bulge.
\begin{figure}[htb] 
\center{\epsfig{file=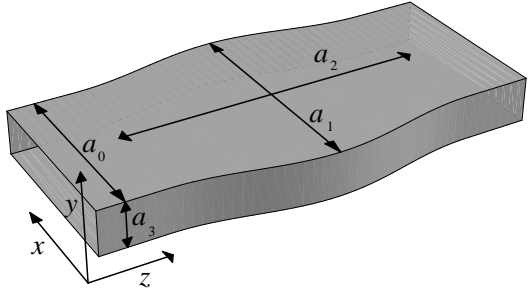,height=40mm,angle=0} }
\caption{Geometry of near perfect wire or ``open quantum dot'',
parametrised as 
$x_{0}(z)=\frac{1}{2}[a_{0}+(a_1-a_0) \cos ^{2}\pi z/a_{2}] $ for
$|z|\leq\frac{1}{2}a_{2}$ and $x_{0}(z)\equiv \frac{1}{2}a_{0}$
otherwise. }
\end{figure} 

To be definite, we choose parameters appropriate to GaAs for the wire
and Al$_{\delta}$Ga$_{1-\delta}$As for the barrier with $\delta$ such
that $V_{0}=0.4$ eV, which is close to the crossover to indirect
gap. Band non-parabolicity is neglected and we use the GaAs effective
mass, $m^{\ast }=0.067m_{0}$, neglecting its variation across the
boundary.  The corresponding one-electron Schr\"{o}dinger equation reads
\begin{equation}
-{\frac{\hbar ^{2}}{2m^{\ast }}} \nabla^2 \Psi (x,y,z)+V\Psi
(x,y,z)=(E_0+E)\Psi (x,y,z),  \label{schroedinger}
\end{equation}
\noindent
where $E$ is electron energy measured from the lowest transverse
channel in the straight part of the wire with energy $E_0$ [which is
$V_0$-dependent and 
$E_0=h^2/(8m^*)(a_0^{-2}+a_3^{-2})$ for $V_0\to \infty$].
The wave function is expanded in elementary modes or channels, 
$\Psi \left( x,y,z\right) =\sum _{n}\Phi _{n}
\left( x,z\right) \chi \left( y\right) \psi _{n}\left( z\right)$,
where the basis functions $\Phi _{n}(x,z)$ are orthogonal solutions of
one-dimensional (1D) Schr\"{o}dinger equations in the $x$-direction
for fixed $z$. We choose $a_3 \ll a_0$ and hence only the lowest
$y$-channel solution, $\chi \left( y\right)$, is relevant.

If such a wire is connected to metallic source-drain contacts,
electrons can flow into the wire region.  At least one electron will
become bound in the bulge region of the wire. The number of bound
electrons depends on both the Fermi energy and the relative size of
the bulge (i.e. parameters $a_1$ and $a_2$). The single electron
problem may be reduced to a quasi-1D $N$-component differential
equation \cite{nakazato91,ramsak98}. 
>From the solution of the scattering problem, the
conductance is calculated from the usual Landauer-B\"{u}ttiker
formula.  For $a_1/a_0 \gg 1$ many channels are needed while for $a_1
\sim a_0$ the inter-channel mixing can be neglected and the
conductance is very similar to that of a perfect straight wire with
conductance steps in multiples of $G_0$. For energies $E<0$, the
solutions of Eq.~(\ref{schroedinger}) are bound states.
          
We consider the interacting electron problem with wire
parameters in a range which ensures that only one electron occupies a
bound state and that restriction to a single channel near the
conduction edge is an excellent approximation. This is the case when
the bulge is sufficiently weak with $a_1 {> \atop \sim} a_0$.
The electron wave function is then determined from a 1D Schr\" odinger
equation for $\psi_0(z)$ as in Ref.~\cite{ramsak98}. Within the
single-channel approximation we consider the interacting two-electron
problem in which one electron is bound in the quantum dot region.
It should be noted that the existence of a
single-electron bound state is guaranteed in 1D and in this sense is a
universal feature. With the chosen parameter range, a second electron
cannot be bound due to the effective 1D Coulomb repulsion $U(z,z')$
between the electrons,
\begin{equation}
\fl U\left( z,z^{\prime }\right) =\frac{e^{2}}{4\pi \varepsilon 
\varepsilon _{0}}\int \!\!\!\!\int\!\!\!\!
\int\!\!\!\! \int \frac{\left| \Phi _{0}\left( x,z\right) 
\Phi _{0}\left( x^{\prime },z^{\prime
}\right) \chi \left
( y\right)  \chi \left( y^{\prime }\right) \right| ^{2}
\textrm{d}x\textrm{d}
x^{\prime }\textrm{d}y\textrm{d}y^{\prime }}
{\sqrt{\left( x-x^{\prime }\right) ^
{2}+\left( y-y^{\prime }\right) ^{2}+\left( z-z^{\prime }\right)
^{2}}}.\label{u}
\end{equation}
The dielectric constant is taken as $\varepsilon=12.5$, appropriate
for GaAs. We solve the two-electron scattering problem exactly
subject to the boundary condition that the asymptotic states consist
of one bound electron in the ground state and one free electron.

In Fig.~2(a) we show plots at zero temperature of ${\cal T}_{{\rm
s}}(E)$ and ${\cal T}_{t}(E)$ for a typical wire with the geometry of
Fig.~1. The thin dotted line represents the non-interacting result,
independent of spin.  In Fig.~2(b) the conductance $G(E)/G_0$ is shown,
as calculated from Eq.~(\ref{landauerT}) for various temperatures.
The resonances have a strong temperature dependence
and, in particular, the sharper singlet resonance is more readily
washed out at finite temperatures. However, it should be noted that
resonances survive to relatively high temperatures, because the width
of the wire, which dictates the energy scale, is small ($a_0=10$~nm)
\cite{ramsak98}.  Note that for weak coupling, the energy scale is set
by the $x$-energy of the lowest channel, $\thicksim a_{0}^{-2}$ and
hence the conductance vs $Ea_{0}^{2}$ with $U a_0$ fixed is roughly
independent of $a_{0}$ (the scaling would be exact for
$V_0\to\infty$).
\begin{figure}[htb]
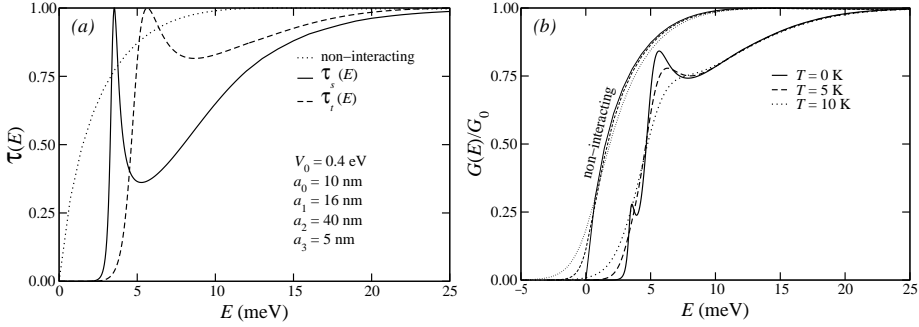

\center{\epsfig{file=Fig2a.eps,height=60mm,angle=-90}
\epsfig{file=Fig2b.eps,height=60mm,angle=-90}}
\caption{(a) Zero temperature singlet transmission probability ${\cal
T}_{{s}}(E)$, and triplet ${\cal T}_{t}(E)$ with full and dashed
lines, respectively.  The energy $E$ is measured from the lowest
transverse channel in the straight part of the wire. The dotted line
represents the corresponding non-interacting result.  (b) Total
conductance, $G(E)$, for the temperature range
$T\leq10$~K, where resonances are still discernible. The thin lines
show the corresponding non-interacting result.}
\end{figure}

The effect of elevated temperatures is mainly to smear the resonances.
The effect of a magnetic field on conductance is much more subtle
\cite{thomas96,ramvall97} and a complete general theory is not
presently available.  For the special case of in-plane magnetic field
(parallel to the $x$-$z$ plane), however, an estimate can be obtained
as follows.
We assume that the bound electron in the initial state is polarised
with spin $\downarrow$.  This assumes that
the bound electron will reach its lowest Zeeman state
between scattering events, whereas the effect of the field on the
electrons in the leads near the Fermi energy will be to simply change
their densities of states, as in Pauli paramagnetism. Thus the energy
of the localised electron will be lowered by $E_B=\frac{1}{2} g^*\mu_B
B$ whereas the electron densities will be
$\rho_\uparrow(E,B)=\rho_\uparrow(E-E_B,0)$ and
$\rho_\downarrow(E,B)=\rho_\downarrow(E+E_B,0)$. Although the densities of
up and down spin electrons are no longer equal in finite magnetic
fields, the conductances of each spin species will be independent of
their densities in the Landauer-B\" uttiker formula, due to the usual
cancelation with group velocity. Hence, assuming that the
transmission amplitudes have the same dependence on kinetic energy as
in zero field, the conductance is $G(E,B)=G_0 {\cal T}(E,B)$, where
\begin{eqnarray}
\fl{\cal T}(E,B)=
\frac{1}{2}\left( \left| t_{\downarrow \downarrow \rightarrow \downarrow 
\downarrow }(E+E_B)\right|^{2}+
\left| t_{\uparrow \downarrow \rightarrow    
\uparrow \downarrow }(E-E_B)
\right|^{2}+\left| t_{\uparrow \downarrow    
\rightarrow \downarrow \uparrow }(E-E_B)
\right| ^{2}\right) \nonumber \\
\lo=\frac{1}{2}{\cal T}_{{\rm t}}(E+E_B)+\frac{1}{4}[{\cal T}_{\rm s}(E-E_B)
+{\cal T}_{\rm t}(E-E_B)].
\label{b}
\end{eqnarray}
We have included the spin-flip term in this equation, which assumes
that the scattered electron, which lies $2E_B$ below the Fermi energy,
is not reflected by the collector. This necessitates inelastic
processes in the collector and the approximation may break down in some
circumstances which we shall not consider further here. 

In any case, $t_{\uparrow \downarrow \to \downarrow \uparrow}(E-E_B)=
t_{\uparrow \downarrow \to \uparrow \downarrow}(E-E_B)=0$ when $E\leq
E_B$ for which we get from Eq.~(\ref{b})
\begin{equation}
G(E,B)=\frac{e^2}{h}{\cal T}_{\rm t}(E+E_B).\label{bb}
\end{equation}
This is plotted in Fig.~3(a) for $T=3$~K together
with the corresponding results for non-interacting electrons and a
straight wire. We see that these curves are very similar with a
plateau at $e^2/h$ but with the interacting case displaced to the
right (due to the Coulomb repulsion) and showing a slight dip, due to
the broad triplet resonance. This curve is very similar to high-field
experimental curves on gated 2DEG wires which   
show the ``0.7'' anomaly \cite{thomas98}, further supporting the view that
an electron is weakly bound in the wire.  In Fig.~3(b) $G(E,B)$ for
$T=3$~K is presented for magnetic field increasing from zero in steps
with $\Delta E_B=0.5$~meV and for clarity the curves have been shifted
by $2E_B$ to the right with increasing $E_B$. We present results for
$a_0=10$~nm, but note that $E_B$ also obeys the above mentioned scaling
$E_B a_0^2$ with varying $a_0$. Magnetic fields which would give
substantial effects in e.g.  narrow ``v''-groove wires
\cite{kaufman99}, would have to be very large, since $E_B=1$~meV
corresponds to large $g^*B\sim 35$~T. However, due to ``$E_B a_0^2$''
scaling, the corresponding value for a wider wire with $a_0\sim50$~nm
would be only $\sim 1.4$~T.  Also plotted in Fig.~3(b) for comparison are
the corresponding results for the non-interacting electron case (dotted)
and the perfectly straight wire (dashed), with $E_B=2$~meV. In this
figure we have indicated with a dot the points $E=E_B$. To the left of
these points $G$ satisfies Eq.~(\ref{bb}) whereas at high energies
$t_{\uparrow \downarrow \to \uparrow \downarrow}$ and
$t_{\uparrow \downarrow \to \downarrow \uparrow}$ are non-zero in
Eq.~(\ref{b}). As argued above, these parts of the curves should be
treated with caution though they are expected to be more reliable at
lower fields. 
\begin{figure}[htb]
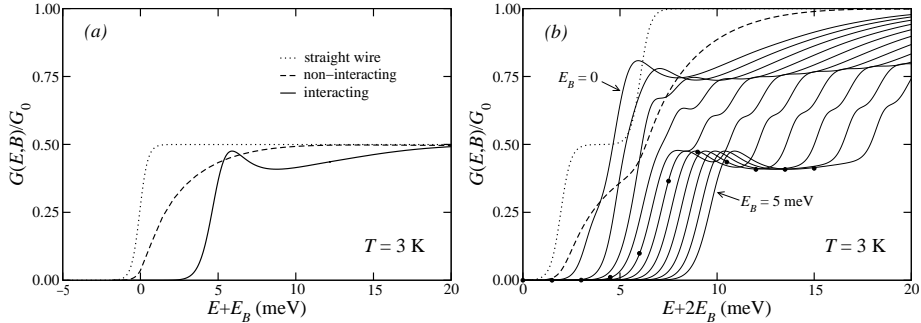
  
\center{\epsfig{file=Fig3a.eps,height=60mm,angle=-90}
\epsfig{file=Fig3b.eps,height=60mm,angle=-90}}
\caption{(a) $G(E,B)$ in large magnetic field, Eq.~(\ref{bb}), shown
together with non-interacting result (dashed) and corresponding result
for a perfectly straight wire (dotted).  (b) $G(E,B)$ for $T=3$~K and the
geometry of the wire shown in Fig.~2.
Successive traces represent results for $E_B$ incremented, in steps
$\Delta E_B=0.5$~meV and for clarity have been horizontally offset by
$2E_B$. Also shown is the non-interacting result (dashed) and perfect
straight wire result (dotted) for $E_B=2$~meV.}
\end{figure}
              
In summary, we have shown that quantum wires with weak longitudinal
confinement, or open quantum dots, can give rise to spin-dependent,
Coulomb blockade resonances when a single electron is bound in the
confined region. This is a universal effect in one-dimensional systems
with very weak longitudinal confinement. The emergence of a specific
structure at $G(E)\sim \frac{1}{4}G_{0}$ and $G\sim
\frac{3}{4}G_{0}$ is a consequence of the singlet and triplet 
nature of the resonances and the probability ratio 1:3 for singlet and
triplet scattering and as such is a universal effect. A comprehensive numerical
investigation of open quantum dots using a wide range of parameters
shows that singlet resonances are always at lower energies than the
triplets, in accordance with the corresponding theorem
for bound states \cite{lieb62}.
With increasing in-plane magnetic field, the resonances
shift their position and a plateau $G(E)\sim e^2/h$
emerges. The effect of a magnetic field is observable only in 
relatively wider quantum wires, due to the intrinsic energy
scale $\propto a_0^{-2}$.

Finally, we speculate on how these results might change if more than
one electron is confined longitudinally in the wire. This possibility
could arise, for example, in long, near perfect wires with a long weak
confinement region. Theoretically this becomes a more complicated
spin-dependent scattering problem \cite{landauQM}. The conductance
would then again show resonance anomalies with positions determined by
the weights of the spin states. The generalised Landauer-B\" uttiker
formula for such cases was discussed recently by Flambaum and Kuchiev
\cite{flambaum99}, who also derived independently the formula for the
singlet/triplet case discussed above and in Ref.~\cite{rejec99}.  In
the case when two electrons are bound, i.e. a conduction electron
scattering from two bound electrons in the confinement region, the
relevant resonant states of three electrons will be doublets and
quartets. When the length of the longitudinal confinement region is
somewhat greater than a Bohr radius, we are in the quasi-1D strong
correlation regime for which we expect a low-lying manifold of spin
states, well separated from higher-lying states and described by a
Heisenberg model, as in a 1D quantum dot with 3 electrons. This spin
manifold contains $2^3=8$ states which split into two doublets and a
quartet and we expect a doublet to be lowest by the Lieb-Mattis
theorem. This is consistent with the exchange being antiferromagnetic,
which is the case in a truly closed 1D quantum dot with 3 electrons,
for which the quartet is highest in energy \cite{JHJ}.  If this
picture holds for the resonant bound states, then we should get two
doublet resonances with weight $\frac{1}{4}$ each, followed by the
quartet at highest energy with weight $\frac{1}{2}$.  This latter
resonance will give a conductance anomaly near $e^2/h$. Furthermore,
this resonance will be broader than the doublet resonances since it is
somewhat higher in energy than the two singlets and is thus expected
to be more pronounced at finite temperatures. Conductance anomalies
close to $e^2/h$ have been observed very recently in long, clean and
nominally straight wires \cite{kane98}. 
This is consistent with the above scenario though we must await
detailed calculations in this strong correlation regime for a more
complete picture.
 
The authors wish to acknowledge A.V. Khaetskii and C.J. Lambert for
helpful comments. This work was part-funded by the U.K. Ministry of
Defence and the EU.


\section*{References}
\begin{thebibliography}{99}
\bibitem{wees88}  {B.J. van Wees {\em et al.}, Phys. Rev. Lett. {\bf 60}, 
848 (1988).} 
\bibitem{wharam88}  {D.A. Wharam {\em et al.}, J. Phys. C {\bf 21}, L209 
(1988).} 
\bibitem{yacoby96}  {A. Yacoby {\em et al.}, Phys. Rev. Lett. {\bf 77}, 4612 
(1996).} 
\bibitem{ramvall97}  {P. Ramvall {\em et al.}, Appl. Phys. Lett. {\bf 71}, 
918 (1997).} 
\bibitem{walther92}  {M. Walther, E. Kapon, D.M. Hwang, E. Colas, and L. 
Nunes, Phys. Rev. B {\bf 45}, 6333 (1992); M. Grundmann {\em et al.}, 
Semicond. Sci. Tech. 
{\bf 9}, 1939 (1994); R. Rinaldi {\em et al.}, Phys. Rev. Lett. {\bf 73}, 
2899 (1994).} 
\bibitem{kaufman99}  {D. Kaufman {\em et al.}, Phys. Rev. B {\bf 59}, R10433 
(1999).} 
\bibitem{kane98} {B.E. Kane {\em et al.}, Appl. Phys. Lett. {\bf 72},
3506 (1998); D.J. Reilly, cond-mat/0001174.} 
\bibitem{landauer57}  {R. Landauer, IBM J. Res. Dev. {\bf 1}, 223 (1957); 
{\bf 32}, 306 (1988).} 
\bibitem{buttiker86}  {M. B\"{u}ttiker, Phys. Rev. Lett. {\bf 57},
1761 (1986).}
\bibitem{houten92} {H. van Houten, C.W.J. Beenakker, and B.J. van Wees, in
{\it Semiconductors and Semmimetals},  edited by R.K. Willardson,
A.C. Beer, and E.R. Weber, (Academic Press, 1992).}
\bibitem{thomas96}  {K.J. Thomas {\em et al.}, Phys. Rev. Lett. {\bf 77}, 
135 (1996); Phys. Rev. B {\bf 58}, 4846 (1998); {\bf 59}, 12252 (1999).} 
\bibitem{thomas98} {K.J. Thomas {\em et al.}, Phil. Mag. B {\bf 77}, 
1213 (1998).}   
\bibitem{fasol94}  {G. Fasol and H. Sakaki, Jpn. J. Appl. 
Phys. {\bf 33}, 879 (1994).}     
\bibitem{maslov95} {D.L. Maslov, Phys. Rev. B {\bf 52}, R14368, 
1995}. 
\bibitem{wang98}  {Chuan-Kui Wang and K.-F. Berggren, Phys. Rev. B {\bf 57}, 
4552 (1998).} 
\bibitem{rejec99} {T. Rejec, A. Ram\v sak, and J.H. Jefferson,
to appear in Phys. Rev. B, 15-Nov-2000.} 
\bibitem{landauQM} {L.D. Landau and E.M. Lifshitz, {\it Quantum
Mechanics} (Pergamon Press, Oxford, 1977).}
\bibitem{oppenheimer28} {J.R. Oppenheimer, Phys. Rev. {\bf 32}, 361 
(1928); N.F. Mott, Proc. Roy. Soc. A {\bf 126}, 259 (1930).} 
\bibitem{bagwell89}  {P.F. Bagwell and T.P. Orlando, Phys. Rev. B
{\bf 40}, 1456 (1989).} 
\bibitem{nakazato91} {K. Nakazato and B.J. Blaikie, J. Phys.: Condens.
Matter {\bf 3} 5729 (1991).}
\bibitem{ramsak98}  {A. Ram\v sak, T. Rejec, and J.H. Jefferson, Phys. Rev. 
B {\bf 58}, 4014 (1998).}
\bibitem{lieb62} {E. Lieb and D. Mattis, Phys. Rev. {\bf 125}, 164
(1962).} 
\bibitem{flambaum99} {V.V. Flambaum and M.Yu. Kuchiev, cond-mat/9910415.}
\bibitem{JHJ} {J.H. Jefferson and W. H\"{a}usler, Phys. Rev. B {\bf
54}, 4936 (1996).} 
\end{thebibliography}
\end{document}